\documentclass[twocolumn,showpacs,preprintnumbers,amsmath,amssymb]{revtex4}
\usepackage{graphicx}% Include figure files
\usepackage{dcolumn}% Align table columns on decimal point
\usepackage{bm}% bol\normalfontd math
\usepackage{epsfig}
\usepackage{color}
\begin{document}

%\preprint{APS/123-QED}

\title{Eshelby description of highly viscous flow III}% Force line breaks with \\

\author{U. Buchenau}
 \email{buchenau-juelich@t-online.de}
\affiliation{%
Forschungszentrum J\"ulich GmbH, J\"ulich Centre for Neutron Science (JCNS-1) and Institute for Complex Systems (ICS-1),  52425 J\"ulich, GERMANY
}%

\date{September 27, 2018}% It is always \today, today,

\begin{abstract}
The recent Eshelby description of the highly viscous flow leads to the prediction of a factor of two different viscosities in stationary and alternating flow, in agreement with experimental evidence. The Kohlrausch barrier density increase with increasing barrier height finds a physical justification in the Adam-Gibbs increase of the number of structural alternatives of the Eshelby region with its increasing size. The new Ansatz allows to determine the number of atoms or molecules in the rearranging Eshelby domains from a combination of dynamic shear relaxation and calorimetric data.
\end{abstract}

%\keywords{Suggested keywords}%Use showkeys class option if keyword                              %display desired
\maketitle

\section{Introduction}

The highly viscous flow of undercooled liquids close to the glass transition is still a controversial topic  \cite{bnap,greg,nemilov,angell,dyre,cavagna,tinaa,bb,ngai,shov2015,royall,lerner}.

Though the flow has a sharp well defined terminal relaxation time $\tau_c$ \cite{greg,tinaa}, on the fast side it stretches over many relaxation time decades, roughly describable in terms of a Kohlrausch function with Kohlrausch parameters $\beta$ around one half \cite{bnap,albena}. But there is not yet a generally acknowledged theoretical explanation for the stretching.

A promising approach to understand the highly viscous flow in terms of reversible and irreversible Eshelby transitions has been developed in Paper I \cite{asyth} and Paper II \cite{asyth1} of this series. An Eshelby transition is a structural rearrangement of a region of $N$ atoms or molecules in the undercooled liquid, which changes the shape of the region and, consequently, changes the elastic misfit of the region to the viscoelastic surroundings \cite{eshelby}.

Calculating the viscosity from the shear strain fluctuations, Paper I of this series showed that the terminal relaxation time $\tau_c$ is a factor of eight longer than the Maxwell time $\tau_M=\eta/G$ ($\eta$ viscosity, $G$ short time shear modulus), and derived a well-defined relaxation time distribution for the irreversible structural relaxation processes.

The exact results of Paper I \cite{asyth} are limited to the irreversible processes. Paper II \cite{asyth1} extended this limited theoretical access to a model description of irreversible and reversible relaxation processes by a Kohlrausch Ansatz. Though still a model with one adaptable free parameter (the Kohlrausch $\beta$), the  description allows for satisfactory descriptions of experimental data. 

Here, Paper III shows that the stationary flow viscosity is a factor of two larger than the one calculated from the shear strain fluctuations, which one measures under alternating shear stress. In addition, the paper justifies the Kohlrausch Ansatz, derives a basic relation between the Kohlrausch $\beta$, the structural entropy and the number $N_c$ of particles of the Eshelby region at $\tau_c$, and discusses possible deviations from this relation.

After this introduction, Paper III derives the relation between the two different viscosities and presents the Adam-Gibbs explanation of the Kohlrausch time dependence in Section II. Section III applies the scheme to experimental shear data in several glass formers. Section IV discusses the results and draws some conclusions.

\section{Eshelby description of highly viscous flow}

\subsection{The two viscosities}

Paper I derived the viscosity $\eta_{fluct}$ from the shear strain fluctuations of the irreversible Eshelby transitions \cite{asyth}, providing the relation between the Maxwell time $\eta_{fluct}/G$ and the Eshelby region lifetime $\tau_c$
\begin{equation}\label{etafluct}
	\tau_c=8\frac{\eta_{fluct}}{G},
\end{equation}
where $G$ is the short time shear modulus. Paper II \cite{asyth1} has shown that this relation holds indeed within experimental error in measurements with an alternating shear strain.

But the relation does not hold for the viscosity $\eta_s$ of the stationary flow. An irreversible Eshelby transition changes the elastic shear misfit of the transforming region. This irreversible change does not only change the shear state of the region itself, but also the one of an equivalent volume in the surroundings \cite{eshelby}. For a given volume in space, this implies that its shear state changes twice in an irreversible way during the lifetime $\tau_c$, once by direct jumps which involve the volume and once by jumps in the neighborhood. 

In stationary flow, the volume is recreated twice during $\tau_c$, each time in equilibrium with the actual shear state at the given time. This continuous creation process implies a back-lag of the average equilibrium shear position $\epsilon_{eq}$ of the state by $\sigma\tau_c/4\eta_s$, where now $\eta_s$ is the viscosity of the stationary flow. Since the average equilibrium shear position is given by $\sigma/G$, one concludes
\begin{equation}\label{etastat}
	\tau_c=4\frac{\eta_s}{G},
\end{equation}
from which follows that the viscosity $\eta_s$ of the stationary flow is a factor of two larger than the viscosity $\eta_{fluct}$ measured in an alternating shear stress field.

The physical point of this consideration is that the stationary viscosity $\eta_s$ does only count the number of times in which the volume forgets its original shear state. It is important that the original shear strain goes to zero, but the width of the resulting distribution around zero is unimportant. The fluctuation-dissipation theorem does not apply, because in stationary flow the sample is not in thermal equilibrium.

It is difficult to believe that this factor two has escaped the attention of the rheological community for many decades, but Section III A will present experimental evidence for the validity of eq. (\ref{etastat}). The validity of eq. (\ref{etafluct}) has been demonstrated for many examples in Paper II \cite{asyth1}.  

\subsection{Adam-Gibbs interpretation of the Kohlrausch $\beta$}

According to the Adam-Gibbs concept \cite{adam}, the fragility (the dramatic increase of the viscosity of undercooled liquids on cooling) results from the growth of the size of the rearranging regions with decreasing temperature. Here, using the Eshelby region results of Paper I and II, the Adam-Gibbs concept is extended into a full quantitative description of the highly viscous flow, relating the Kohlrausch exponent $\beta$ to the structural excess entropy $S_1$ per atom or molecule.  

For a thermally activated structural rearrangement over an energy barrier $V$, the relaxation time $\tau$ is
\begin{equation}\label{arrh}
	\tau=\tau_0\exp(V/k_BT),
\end{equation}
with $\tau_0=10^{-13}$ s. The relation defines the barrier $V_c$ for the terminal relaxation time $\tau_c$.

In the following, let us describe the relaxation time dependence in terms of the barrier variable $v$
\begin{equation}\label{v}
	v=\frac{V-V_c}{k_BT}=\ln(\tau/\tau_c)
\end{equation}
which is zero at the terminal relaxation time $\tau_c$, with the corresponding energy barrier $V_c$.

For the shear relaxation, a Kohlrausch $\beta$ close to 1/2 means that the shear compliance function $l(v)$ (defined such that $l(v)dv$ gives the contribution to the shear compliance for relaxations between $v$ and $v+dv$ in units of $1/G$, where $G$ is the short time shear modulus), has the form
\begin{equation}
	l(v)=l_0\exp(\beta v),
\end{equation}
with a prefactor $l_0$. If the barrier increase is due to the increase of the number of atoms or molecules in the Eshelby region, $\beta$ must result from the concomitant increase in the number of structural possibilities. 

Let us begin with the simple approximation that the barrier $V_N=NV_1$ for the rearrangement of an Eshelby domain is proportional to the number $N$ of particles. For an Eshelby region consisting of $N$ atoms or molecules, the number $n_s$ of possible stable structures in thermal equilibrium is given by
\begin{equation}\label{ns}
	n_s=\exp(NS_1/k_B),
\end{equation}
where $S_1$ is the structural entropy per particle.

In Section II. C, it will be shown (or at least will be made plausible) that the coupling factor of an Eshelby transformation to an external shear stress does not depend on the size of the Eshelby region. If this is true, the $v$-dependence of $l(v)$ is exclusively due to the density of alternative structures of the Eshelby region with a size corresponding to $v$. Adding one particle to the region increases the structural entropy by $S_1$ and increases the barrier by $V_1$. Therefore
\begin{equation}\label{beta}
	\beta=\frac{S_1T}{V_1},
\end{equation}
a relation which was already given as eq. (19) in Paper II \cite{asyth1}.

With the relation, one can determine the number $N$ of particles of the Eshelby region at a given relaxation time $\tau$, which determines a barrier $V$ via eq. (\ref{arrh})
\begin{equation}\label{N}
	N=\frac{\beta V}{S_1T},
\end{equation}
provided the total barrier is indeed given by $V=V_N=NV_1$. In Section IV, possible deviations from this relation will be discussed.

At the terminal relaxation time $\tau_c$, corresponding to the barrier variable $v=0$ according to its definition in eq. (\ref{v}), the Eshelby rearrangements change from reversible ones for $v<0$ to irreversible ones for $v>0$. For simplicity, the crossover at $v=0$ is taken to be sharp. The contribution of the Eshelby domains above $\tau_c$ to the viscous flow increases with $l_0\exp(\beta v)$, because their density increases, but decreases with their jump rate $1/\tau$. Integrating over all contributions
\begin{equation}
	8=l_0\int_0^\infty\exp((1-\beta)v)dv=\frac{l_0}{1-\beta},
\end{equation}
because the viscous flow amounts to $8/G$ within the time $\tau_c$, which is eight times the Maxwell time $\eta_{fluct}/G$, and $l(v)dv$ gives the compliance contribution in units $1/G$.  The relation fixes the prefactor $l_0=8(1-\beta)$.

As shown in Paper II, the reversible Eshelby transitions have two weakening factors: a factor of 1/2 because they do not relax the surroundings, and a factor $f_{r0}=0.4409$ from the different effect of the asymmetries on reversible and irreversible transitions \cite{asyth1}. This implies a density of reversible compliance processes
\begin{equation}\label{lv}
	l_r(v)=4f_{r0}(1-\beta)\exp(\beta v)\exp(-\exp(v)),
\end{equation}
where the last factor $\exp(-\exp(v))$ ensures the upper cutoff at $v=0$.

Equation (\ref{lv}) corrects eq. (13) of Paper II, which had the slightly different prefactor $f_{r0}(8-4\beta)/3$ (both are equal at $\beta=1/2$). The correction is so small that the fitted values in Table I and Table II of Paper II remain the same within the error bars \cite{asyth1}. 

Integration of $l_r(v)$ over $v$ provides a total relaxational contribution to the reversible compliance of $4f_{r0}(1-\beta)/\beta$ in units of $1/G$. To this, one has to add $1/G$ from the elastic shear response, so the total recoverable compliance $J_0$ is given by
\begin{equation}\label{gj0}
	GJ_0=1+1.7636\frac{1-\beta}{\beta}.
\end{equation}

\subsection{Coupling factors}

A structural rearrangement changes the shape of the region, changing the elastic shear misfit of the region to the viscoelastic surroundings. This implies that the transition is an Eshelby transition \cite{eshelby}. The question to be answered is: How does the average elastic shear misfit change depend on the size of the region?

In order to answer this question, one has to decompose the structural rearrangement of many atoms or molecules into local rearrangements, small enough to define a uniform local shear strain change. The total shear change of the region is obtained by summing up all uniform local changes.

For simple liquids consisting of single atoms, a uniform local strain is defined by the rearrangement motion of three atoms which are nearest neighbors to each other, either before or after the rearrangement. Their motion in the rearrangement can be decomposed into a translation, a rotation, an expansion and a local uniform shear.

It is possible that the Eshelby transition in simple liquids consists essentially of transformations of nearest neighbors to second nearest neighbors and viceversa. This is consistent with the finding of a string motion with amplitudes of about ten percent of the atomic distance, discovered first in low-barrier relaxations in glasses \cite{olig} and later in undercooled liquids \cite{glotzer}. For two atoms undergoing this nearest to second-nearest neighbor transformation, it should almost always be possible to find a third atom which stays nearest neighbor to the two transforming ones. The transformation of the three atoms from an equilateral triangle to a right isosceles triangle causes a local shear distortion $\epsilon_1$. Taking the extension of the atomic spheres into account, the equilateral triangle lengthens its basis from 2 atomic distances to $\sqrt(2)+1$ and shortens its height from $1+\sqrt{3}/2$ to $1+\sqrt{1/2}$, so $\epsilon_1=0.292$.

For the whole Eshelby region with $N$ atoms, this local shear strain contributes $N_1\epsilon_1/N$, where in this case $N_1=3$, to the total shear strain change in the rearrangement.

An alternative is the gliding triangle transformation of the six atoms around an octahedral hole to a double tetrahedron, with the two tetrahedra connected at the tops \cite{buscho}. Formulating it in this way, one realizes that the gliding triangle transformation creates and annihilates octahedral holes in a simple undercooled liquid. In this case, $N_1=6$ and $\epsilon_1^2=1/2$.

If the uniform local shear strain changes are uncorrelated, one has to sum up the squares of their contributions to obtain the square of the elastic misfit change of the Eshelby region. The result is inversely proportional to $N$
\begin{equation}\label{en}
	\overline{\epsilon_N^2}=\frac{N_1\epsilon_1^2}{N}.
\end{equation}
Here $N_1$ is the number of atoms or molecules in the uniform-strain subregions and $\epsilon_1^2$ is the average squared expectation value of the uniform shear strain, provided the proposed mechanism is dominating and space-filling.

For rigid molecules, the uniform-strain subregion consists of two neighboring molecules, $N_1=2$. Their motion in the structural rearrangement can be decomposed into three joint translations, three joint rotations, and six motions against each other, which are describable as one expansion and five uniform shear strains.

The case of flexible molecules is also accessible to a similar analysis, if one can specify the possible molecular conformations. An example is the Helfand crankshaft in polymers \cite{helfand}, for which the shear distortion has been derived in Paper II.

The whole sample with $N_{tot}$ particles has the squared shear fluctuation $\epsilon_{tot}^2$, the sum of the Eshelby contributions of eq. (\ref{en}) multiplied with $N^2/N_{tot}^2$. Since the sample has $N_{tot}/N$ Eshelby domains with $N$ particles, the factor $1/N$ of eq. (\ref{en}) cancels out. As a consequence, the size dependence of the Eshelby contribution to the shear compliance is exclusively dominated by the increase of the number of structural possibilities.

\subsection{Eshelby density in five-dimensional shear space}

The occurrence of the highly viscous flow requires a certain density of Eshelby states at the barrier $V_c$ belonging to the terminal relaxation time $\tau_c$ according to the Arrhenius equation (\ref{arrh}). In order to calculate this density, it is necessary to reformulate and slightly correct the derivation of paper I.  

Consider a structural jump of an Eshelby region over a barrier larger than $V_c$ into another structure with a different elastic misfit to the surroundings. Before the jump, the region has a shear misfit angle $\epsilon_0$ (in radian) with respect to the surrounding viscoelastic matrix.  The region jumps into another shear misfit $\epsilon$. 

According to the Eshelby theory \cite{eshelby}, the shear energy increase or decrease by the jump is given by
\begin{equation}\label{delta0}
	\Delta=\frac{GNV_a\epsilon^2}{4}-\frac{GNV_a\epsilon_0^2}{4}.
\end{equation}
Here $NV_a$ is the volume of the region consisting of $N$ particles and $G$ is the short time shear modulus. Half of each of the two distortion energies is shear energy of the region, the other half is shear energy of the surroundings.

Let us define the shear states $\epsilon_0$ and $\epsilon$ by the dimensionless quantities $e_0$ and $e$ with
\begin{equation}\label{e0e}
	e_0^2=\frac{GNV_a\epsilon_0^2}{4k_BT}\ \ \ \ e^2=\frac{GNV_a\epsilon^2}{4k_BT},
\end{equation}
where $V_a$ is the volume of the atom or the molecule.

From the point of view of elasticity theory, the surroundings of the region react at short times like an isotropic elastic medium, describable by a strain tensor with one compression and five independent shear components. The shear misfits $e_0$ and $e$ are thus vectors in a five-dimensional shear misfit space.

The shear energy $E_s$ of the saddle point, supposed to lie in the middle between the two structures, is
\begin{equation}
	\frac{E_s}{k_BT}=\frac{1}{2}(e_0^2+e^2+2\vec{e_0}\cdot\vec{e}).
\end{equation}

The calculation of the escape rate from the state $e_0$ requires an integral over all possible $e$-values. In this integral, the contribution of the scalar product cancels. Therefore the effective barrier between $e_0$ and $e$ is changed in units of $k_BT$ by the amount $(e_0^2+e^2)/2-e_0^2=(e^2-e_0^2)/2$. Thus the jump rate from $e_0$ to $e$ gets a factor $\exp((e_0^2-e^2)/2)$ from the difference in the shear misfits.

With this, the state $e_0$ has the escape rate
\begin{align}\label{e0r}
	r=N_sr_V\frac{8\pi^2}{3}\int_0^\infty\exp((e_0^2-e^2)/2)e^4de \nonumber\\
	=4\sqrt{2}\pi^{3/2}N_sr_V\exp(e_0^2/2),
\end{align}
where $N_s$ is the density of stable structural alternatives to the initial state in the five-dimensional $e$-space and $r_V$ is the jump rate for the barrier height $V$ between two states with the same elastic misfit energy.

In thermal equilibrium, the states $e$ in the five-dimensional shear misfit space have an average energy of 5/2 $k_BT$ in the normalized distribution
\begin{equation}
	p(e)=\frac{1}{\pi^{5/2}}e^4\exp(-e^2).
\end{equation}
The prefactor corrects the one of eq. (3) in Paper I \cite{asyth}, a mistake which does not invalidate the results of Paper I.

The average escape rate in thermal equilibrium is
\begin{equation}\label{rav}
	4\sqrt{2}\pi^{3/2}r_VN_s\int_0^\infty\exp(e^2/2)p(e)de=\frac{12N_sr_V}{\pi}.
\end{equation}

At the terminal relaxation time $\tau_c$, $V=V_c$ and the rate is $r_V=1/2\tau_c$ for a symmetric double well potential, in which both jump directions contribute to the relaxation. All barriers with $V>V_c$ contribute to the highly viscous decay, with a contribution which increases with $\exp(\beta v)$ from the density and decreases with $\exp(-v)$ from the decreasing escape rate. This leads to an average escape rate
\begin{equation}
	\overline{r_V}=\frac{1-\beta}{(2-\beta)2\tau_c},
\end{equation}
for $\beta=1/2$ to $\overline{r_V}=1/6\tau_c$. Inserting this value into eq. (\ref{rav}) and demanding the escape rate $1/\tau_c$, one sees that one needs the density $N_s=\pi/2$, a value close to 1, to keep the highly viscous flow going.

The relation between $N_s$, the density of structurally stable states in the five-dimensional $e$-space, and the total number $n_s$ of structurally stable states of a given Eshelby region, is provided by eq. (\ref{en}). In terms of the definition of $e$ in eq. (\ref{e0e}), eq. (\ref{en}) defines a mean square $\overline{e^2}$
\begin{equation}\label{eq}
	\overline{e^2}=\frac{N_1GV_a\epsilon_1^2}{4k_BT_g}.
\end{equation}

It follows that the $n_s$ states of the Eshelby region have a gaussian distribution in $e$-space
\begin{equation}\label{nsNs}
	\frac{n_s}{N_s}=\frac{8\pi^2}{3}\int_0^\infty\exp(-e^2/2\overline{e^2})e^4de=4\sqrt{2}\pi^{5/2}\overline{e^2}^{5/2}.
\end{equation}

\section{Comparison to experiment}

\subsection{The two viscosities}

Section II A compared two theoretical relations, eq. (\ref{etafluct}) for the viscosity $\eta_{fluct}$ in an alternating field and eq. (\ref{etastat}) for the stationary flow viscosity $\eta_s$, concluding that $\eta_s$ should be a factor of two larger than $\eta_{fluct}$.

If one looks for direct experimental evidence, a comparison of stationary and alternating measurements of the same sample at the same temperature, one does not find anything in the literature. Obviously, everybody just assumed the two viscosities to be equal, attributing differences between alternating and stationary measurements to different temperature calibrations (remember that the viscosity is strongly temperature-dependent).

%%%%%%%%%%%%%%%%%%%%% begin figure %%%%%%%%%%%%%%%%%%%%%%%%%%%%%%%%%%%%%
\begin{figure}   
\hspace{-0cm} \vspace{0cm} \epsfig{file=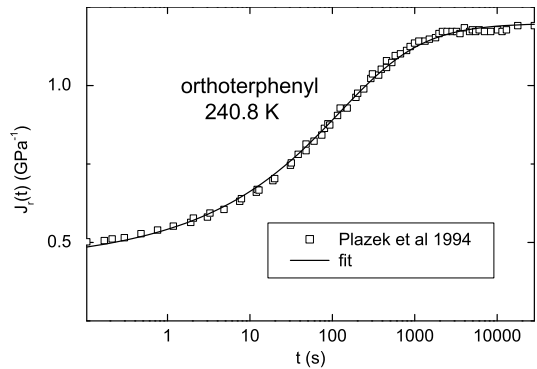,width=7 cm,angle=0} \vspace{0cm} \caption{Recoverable shear relaxation $J_r(t)$ at the glass temperature of orthoterphenyl \cite{bero}, together with a fit in terms of eq. (\ref{gjr}) with the parameters given in Table I. These parameters are consistent with eq. (\ref{etastat}) and not with eq. (\ref{etafluct}).}
\end{figure}
%%%%%%%%%%%%%%%%%%%%% end figure %%%%%%%%%%%%%%%%%%%%%%%%%%%%%%%%%%%%%%%

But one can look for evidence for the validity of the two equations themselves. For eq. (\ref{etafluct}), the alternating case, ample evidence has been presented \cite{asyth1} in Paper II. It remains to show that eq. (\ref{etastat}) for the stationary viscosity $\eta_s$ is also valid.

The best experiments for this purpose are those by Plazek and collaborators \cite{tnb,bero,tnb2,se}, measuring first the viscosity of a more or less cylindrical sample under constant shear stress, and then the recoverable $t^\beta$-response $J_r(t)$ from the reversible Eshelby transitions after removal of the shear stress, a measurement which takes several days. In these experiments, it turned out to be possible to scale data of different temperatures together to a common master curve, using the shift factors of the measured viscosities.  

Fig. 1 shows this master curve of $J_r(t)$ for orthoterphenyl at its glass temperature $T_g=240.8$ K, plotted as a function of time \cite{bero}. The line represents the fit function
\begin{equation}\label{gjr}
	GJ_r(t)=1+(GJ_0-1)\left(\frac{t}{t+\tau_c}\right)^\beta
\end{equation}
with the four parameters short time shear modulus $G$, zero frequency recoverable compliance $J_0$, terminal relaxation time $\tau_c$ and Kohlrausch exponent $\beta$. 

\begin{table}[htbp]
	\centering
		\begin{tabular}{|c|c|c|c|c|c|c|c|}
\hline
substance                             &$T_g$  &$\eta_s$ & $G$  &$GJ_0$&  $\beta$   &  $\tau_c$ &$\tau_cG/\eta_s$  \\
\hline
                                      & K     &GPa s    & GPa  &      &            &    s      &                  \\
\hline
6-PPE                                 & 248.1 &550.8    & 1.29 & 2.2  &  0.49      &  798      &    1.9           \\
aroclor                               & 252.2 &569.2    & 2.22 & 2.3  &  0.44      &  573      &    2.2           \\
OTP                                   & 240.8 &224.1    & 2.29 & 2.5  &  0.44      &  256      &    2.3           \\
TNB                                   & 337.4 &223.9    & 1.04 & 2.5  &  0.47      &  531      &    2.5           \\
TNB-2                                 & 345.6 & 41.3    & 1.25 & 3.4  &  0.35      &  187      &    5.7           \\
\hline                                            
		\end{tabular}
	\caption{Check of the validity of eq. (\ref{etastat}) for measured stationary viscosity $\eta_s$-data. Eq. (\ref{etastat}) predicts a value of 4 in the last column of the Table, compatible with the average value 2.9$\pm$2 of the five measurements. The short time shear modulus $G$ and the terminal relaxation time $\tau_c$ were obtained by fitting eq. (\ref{gjr}) to recoverable compliance data obtained in the same experiment which provided $\eta_s$. 6-PPE is a short polyphenylene consisting of six connected phenylene rings, OTP is orthoterphenyl, TNB is trinaphtyl benzene, TNB-2 trinaphtyl benzene with a slightly different molecular conformation. Data for 6-PPE, aroclor and OTP taken from ref. \cite{bero}, for TNB from ref. \cite{tnb}, for TNB-2 from ref. \cite{tnb2}.}
	\label{tab:rse1}
\end{table}

Table I shows the results of the fitting procedure for data \cite{tnb,bero,tnb2} of five molecular glass formers. The experimental uncertainty for $\tau_c$ is rather large, a factor of two, which results from the freedom provided by the other three parameters. Nevertheless, the average value of the five measurements, $\tau_cG/\eta_s=2.9\pm$2, is even smaller than the value of 4 predicted by eq. (\ref{etastat}). The relaxational part of the recoverable compliance, $GJ_0-1$, is in all five cases only about 2/3 of the prediction of eq. (\ref{gj0}).

In the fits of measurements of $G(\omega)$ of Paper II \cite{asyth1}, eq. (\ref{etafluct}) was assumed to be valid. If one repeats them, leaving the ratio as a free parameter, one finds only few measurements which fix the ratio accurately. The viscosity is fixed within a few percent, but the condition $\omega\tau_c=1$ occurs in a region where the viscous response is much larger than the reversible one. Of the measurements in Table I of Paper II, only the ones in the vacuum pump oils DC704 \cite{tina} with $\tau_cG/\eta_{fluct}=8.5\pm$2 and 5-PPE \cite{niss} with $\tau_cG/\eta_{fluct}=10\pm$3 allow an accurate determination, though there are several others which allow to exclude a $\tau_cG/\eta_{fluct}$ which is smaller than 5.

\subsection{Eshelby region sizes}

Let us continue by estimating the sizes of the Eshelby regions responsible for the terminal relaxation at the glass temperature $T_g$. Let us define $\tau_c(T_g)=100$ s. Then from the Arrhenius equation (\ref{arrh}) $V_c=34.5\ k_BT_g$.

To calculate the number $N_c$ of the particles corresponding to $V_c$ from eq. (\ref{N}), one needs the structural excess entropy $S_1$ per particle at the glass temperature. For a rough estimate, one can identify $S_1$ with the full excess entropy of the undercooled liquid over the crystal phase, keeping in mind that the real $N_c$ can be twenty to thirty percent higher because of the neglect of the vibrational part of the excess entropy \cite{se}.

In principle, the determination of $S_1$ requires dedicated heat capacity measurements of both phases, undercooled liquid and crystal. However, if one has only heat capacity measurements of the undercooled liquid at $T_g$, together with a Vogel-Fulcher temperature $T_{VF}$ from dynamical measurements, one can make use of the Adam-Gibbs relation in the form proposed by Richert and Angell \cite{ra}
\begin{equation}
	S_{exc}=S_\infty\left(1-\frac{T_{VF}}{T}\right)
\end{equation}
which has been found to describe the excess entropy of several molecular glass formers fairly well, at least close to $T_g$.

If this holds, one can determine $S_{exc}$ at $T_g$ from the heat capacity difference $\Delta c_p$ between undercooled liquid and glass via
\begin{equation}\label{s1}
	S_{exc}(T_g)=\Delta c_p\left(\frac{T_g}{T_{VF}}-1\right).
\end{equation}

Table II lists values of $N_c(T_g)$ for three network, two metallic and six molecular glass formers. The numbers are much smaller for molecular glass formers, in particular for large flexible molecules like the two vacuum pump oils DC704 and PPE, but also for the rigid small molecules propylene carbonate and m-toluidine. The table contains in the last column values for the mean square shear jump width $\overline{e^2}$, calculated from eq. (\ref{nsNs}), to which we will return in the discussion in Section IV.

\begin{table}[htbp]
	\centering
		\begin{tabular}{|c|c|c|c|c|c|c|c|}
\hline
substance                          &$T_g$  &$T_{VF}$ & $\Delta c_p$  &$S_1/k_B$&  $\beta$   &  $N_c$ &$\overline{e^2}$\\
\hline   
                                       &K      & K    &$\frac{\rm J}{\rm mole K}$& &       &             & \\
\hline
networks                               &       &      &       &               &            &      &        \\
(SiO$_2$)$_3$Na$_2$O                   &736    &418   & 4.66  &0.43           &.38         &30.5  &  25.2  \\
(SiO$_2$)$_2$Na$_2$O                   &715    &461   & 6.07  &0.40           &.38         &32.8  &  25.3  \\
selenium                               &305    &      &       &0.43           &.31         &25.6  &  10.8  \\
\hline
metallic                               &       &      &       &               &            &      &        \\
Pd$_{40}$Ni$_{40}$P$_{20}$             &570    &      &       &0.36           &.41         &39.3  &  38.1  \\
vitralloy-1                            &630    &      &       &0.36           &.43         &41.2  &  50.1  \\
\hline
molecular                              &       &      &       &               &            &      &        \\
DC704                                  &211    &175   & 131.8 & 3.24          &.48         &5.1   &  98.6  \\
PPE                                    &247    &207   & 192.7 & 4.47          &.52         &4.0   & 197    \\
m-toluidine                            &186    &154   &  75   & 1.89          &.47         &8.6   &  88.5  \\
PC                                     &156    &133   &  71.9 & 1.51          &.44         &10.0  &  55.8  \\
PG                                     &166    &117   &  61.2 & 3.08          &.50         &5.6   & 131.7  \\
glycerol                               &190    &135   &  81.7 & 4.00          &.53         &4.6   & 209    \\
\hline
		\end{tabular}
	\caption{Number $N_c$ of particles (atoms, in the molecular glasses molecules) in the Eshelby region responsible for the terminal relaxation time at the glass temperature $T_g$, calculated from eq. (\ref{N}). PC propylene carbonate, PG propylene glycol, Vogel-Fulcher temperatures calculated from the fragility \cite{bnap,angell}; Kohlrausch $\beta$-values from Table I of Paper II \cite{asyth1}, selenium from Fig. 4 (b), PC from an unpublished fit of the author; $S_1$ from measurements of liquid and crystal in selenium \cite{chang}, in Pd$_{40}$Ni$_{40}$P$_{20}$ \cite{wildek} and in vitralloy-1 \cite{busch}; for the other substances calculated from eq. (\ref{s1}), using $\Delta c_p$-values: (SiO$_2$)$_3$Na$_2$O and (SiO$_2$)$_2$Na$_2$O ref. \cite{nem}, DC704 ref. \cite{gundermann}, PPE ref. \cite{cpppe}, m-toluidine ref. \cite{cptol}; propylene carbonate ref. \cite{cppc}; propylene glycol and glycerol ref. \cite{wunderlich}.}
	\label{tab:rse2}
\end{table}

\subsection{Evidence in the glass phase}

The numbers $N_c$ for the two metallic glasses in Table II are corroborated by experimental evidence \cite{atzmon} for atomic number oscillations detected in shear relaxation data in the glass phase of the metallic glass Al$_{86.8}$Ni$_{3.7}$Y$_{9.5}$ at room temperature, at a measurement temperature $T_m=295$ K.

Transforming the signal oscillations into barrier density oscillations, the authors \cite{atzmon} find equally spaced maxima with a separation of about $2k_BT_m$ at room temperature. In terms of the glass temperature of the alloy, which according to a differential thermal scanning analysis \cite{alni} on similar alloys lies between 500 and 600 K, this is a spacing close to $k_BT_g$. If this spacing reflects the increase $V_1$ of the barrier due to the addition of one atom to an Eshelby region, $N_c$ at $T_g$ should be around 34.5, close to the two values in Table I.

But the experiment holds even more valuable quantitative information, namely the existence of a lower cutoff of the barrier distribution.

%%%%%%%%%%%%%%%%%%%%% begin figure %%%%%%%%%%%%%%%%%%%%%%%%%%%%%%%%%%%%%
\begin{figure}   
\hspace{-0cm} \vspace{0cm} \epsfig{file=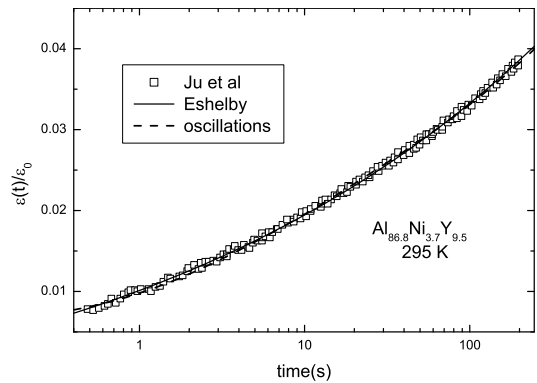,width=7 cm,angle=0} \vspace{0cm} \caption{Shear relaxation under constant stress in a metallic glass at room temperature \cite{atzmon}, together with the oscillatory fit of the authors (dashed line) and the fit with eq. (\ref{etexp}) (continuous line) described in the text.}
\end{figure}
%%%%%%%%%%%%%%%%%%%%% end figure %%%%%%%%%%%%%%%%%%%%%%%%%%%%%%%%%%%%%%%

Fig. 2 shows the data taken by a cantilever under constant shear stress \cite{atzmon}, fitted in terms of
\begin{equation}\label{etexp}
	\frac{\epsilon(t)}{\epsilon_0}=l_at^\alpha-a_0.
\end{equation}
Here $\epsilon(t)$ is the shear strain as a function of time and $\epsilon_0$ is the elastic response under the applied stress. One finds $l_a=0.02$, $\alpha=0.17$ and $a_0=0.01$. The fit (without oscillations in the density) is not much worse than the one with oscillations \cite{atzmon}, which tells one that looking for oscillations in the shear response of glass formers is not an easy task.

The internal friction measurements in the glass phase are conveniently expressed \cite{burel} in terms of the barrier density $f(V)$, where $f(V)dV=\delta G/G$ describes the reduction of the shear modulus by the relaxations with barriers between $V$ and $V+dV$. As pointed by Gilroy and Phillips \cite{gilroy}, $f(V)$ remains temperature-independent on cooling, because one has to reckon with a constant asymmetry distribution between two stable states. 

Since the total reduction of $G$ by relaxations in the glass phase remains small, one can use $\Delta G/G=\delta JG$, where $\delta J$ is the increase of the shear compliance from the relaxations with barriers between $V$ and $V+dV$ in units of $1/G$, together with the definition of $v$ in eq. (\ref{v}) and finds the Eshelby contribution
\begin{equation}
	f(V)=\frac{f_{r}(1-\beta)}{k_BT_g}\exp(\beta(V-V_c)/k_BT_g).
\end{equation}

Sampling this barrier distribution at the temperature $T_m$ as a function of time, one expects
\begin{equation}\label{etth}
	\frac{\epsilon(t)}{\epsilon_0}=1.7636\frac{1-\beta}{\beta}\left(\frac{t}{\tau_c}\right)^{\beta T_g/T_m},
\end{equation}
where now $\tau_c$ is the Arrhenius relaxation time of the barrier $V_c$ at the measuring temperature $T_m$.

From Table I of Paper II \cite{asyth1}, one sees that $\beta$ in metallic glass formers is close to 0.4, so the $\alpha=0.17$ of eq. (\ref{etexp}) implies a glass temperature of $T_g=0.4T_m/0.17=694$ K. This is decidedly higher than the estimate of 500 to 600 K for $T_g$ in ref. \cite{alni}, but one must remember that the sample was made by the melt spinning technique \cite{atzmon}, at a cooling rate of about 10$^6$ K/s.

In fact, if one takes the experimental prefactor $l_a=0.02$, it is compatible with a theoretical $\tau_c$ of 3 teraseconds at room temperature (about 100000 years), which with the same energy barrier translates to 0.7 ms at 694 K, a reasonable $\tau_c$ at this high cooling rate.

Having shown that the experimental $l_a$ and $\alpha$ do indeed correspond to the expectation for a Kohlrausch density of Eshelby regions, one arrives at the question of the meaning of $a_0=0.01$ in the experimental fit of eq. (\ref{etexp}). Obviously, this means that the short-time relaxational response, at times shorter than the first points in Fig. 2, is smaller than one would expect. The value $a_0=0.01$ is consistent with a cutoff shortly below the first point in Fig. 2, at a barrier which according to the oscillations observed in the experiment \cite{atzmon} corresponds to an Eshelby region containing twelve atoms.

Such a low barrier cutoff in metallic glasses is not completely unexpected. In fact, the theoretical analysis of the gliding triangle shear transformation of six atoms \cite{buscho} predicted a barrier close to zero, because the restoring forces from the surroundings compensate the negative curvature at the top of the barrier. For a combination of several such units with different orientations of their local strain, the restoring force from the surroundings is much smaller and should rapidly become negligible. 

%%%%%%%%%%%%%%%%%%%%% begin figure %%%%%%%%%%%%%%%%%%%%%%%%%%%%%%%%%%%%%
\begin{figure}   
\hspace{-0cm} \vspace{0cm} \epsfig{file=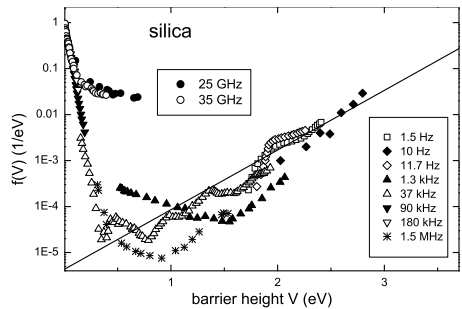,width=6 cm,angle=0} \vspace{0cm} \caption{Comparison of the fitted density of reversible Eshelby states to internal friction data in the glass phase of silica\cite{philmag2002}. The line is calculated from the parameters in Table I of Paper II.}
\end{figure}
%%%%%%%%%%%%%%%%%%%%% end figure %%%%%%%%%%%%%%%%%%%%%%%%%%%%%%%%%%%%%%%

The fact that the Kohlrausch behavior continues down to barriers much lower than $V_c$ has already been shown in Figure 3 of Paper II \cite{asyth1} for silica \cite{philmag2002} and a sodium silicate. In the sodium silicate, the low barrier behavior is clouded by secondary relaxations, but in silica the Kohlrausch density stretches down to barriers of less than half $V_c$. The figure is reproduced here as Fig. 3, because in Paper II it has been inserted with a miscalculated Eshelby line.

But it is also reproduced because one does see marked oscillations of $f(V)$, which are most pronounced in the 37 kHz tensile stress data of Marx and Sivertsen \cite{marx}, with the highest oscillation maximum at 2 eV corroborated by the torsion pendulum 1.5 Hz data of Kirby \cite{kirby}. In the 37 kHz data, one sees four maxima with an average spacing of 0.47 eV, about one ninth of $V_c(T_g)=4.38$ eV. From this, one infers that the maxima correspond to the addition of another SiO$_4$-tetrahedron to the Eshelby domain, so one has about nine formula units, i.e. twenty seven atoms at $V_c$, in good agreement with the values 30.5 and 32.5 calculated for the two sodium silicates in Table II.   

%%%%%%%%%%%%%%%%%%%%% begin figure %%%%%%%%%%%%%%%%%%%%%%%%%%%%%%%%%%%%%
\begin{figure}   
\hspace{-0cm} \vspace{0cm} \epsfig{file=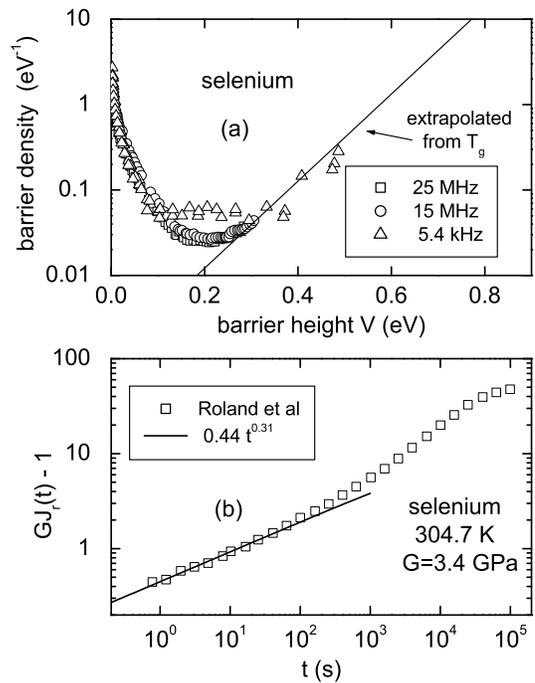,width=7 cm,angle=0} \vspace{0cm} \caption{(a) Barrier density in selenium glass from longitudinal sound absorption \cite{pino} at 25 and 15 MHz and from a paddle oscillator measurement \cite{liu} at 5.4 kHz. The continuous line is the Kohlrausch extrapolation down from the terminal relaxation barrier $V_c$=0.93 eV at the glass temperature 304.7 K, determined from the $J_r(t)$-data \cite{roland} shown in part (b) of the figure, using the shear modulus 3.4 GPa determined independently by two ultrasonic experiments \cite{galli,kozh}.}
\end{figure}
%%%%%%%%%%%%%%%%%%%%% end figure %%%%%%%%%%%%%%%%%%%%%%%%%%%%%%%%%%%%%%%

Fig. 4 shows a third example, selenium. Fig. 4 (a) shows the barrier density calculated from longitudinal ultrasound absorption \cite{pino} using eq. (27) of reference \cite{burel}, in good agreement with the barrier density calculated from a very recent shear experiment at 5.4 kHz with a silicon crystal paddle oscillator covered by a selenium film \cite{liu}. The continuous line is the Kohlrausch expectation extrapolated from the $J_r(t)$-measurements \cite{roland} in Fig. 4 (b). In selenium, one knows the shear modulus $G$=3.4 GPa at 304.7 K from two independent ultrasonic measurements \cite{galli,kozh}, which enables one to plot only the relaxational part $GJ_r(t)-1$ of the recoverable compliance. The upper part shows the steeper rise expected for a polymer \cite{roland}, but the lower part reveals a Kohlrausch $\beta$ of 0.31.

The lowest measured point of the liquid in Fig. 4 (b) corresponds to a barrier height of 0.78 eV, not so far away from the points measured in the glass phase. We conclude that the Eshelby density of the highly viscous flow extends down with an unchanged Kohlrausch $\beta$ to 0.2 eV, about one fifth of the barrier $V_c$ marking the crossover from reversible to irreversible Eshelby transitions at the glass temperature.

\section{Discussion and conclusions}

The theoretical argument of Section II A for the existence of two different viscosities, one in stationary flow and another in alternating fields, is supported by the indirect evidence presented in Section III A. 

But it is clear: what one really needs is a direct proof, a reliable measurement of the two viscosities in the same sample at the same temperature in a dedicated experiment. Such an experiment seems possible within existing rheological technology, and it is very desirable that it should be done.

The prediction of a factor of two between the two viscosities is limited to undercooled liquids. It is no longer valid if the viscous flow is no longer due to thermally activated Eshelby transitions. But even then the two viscosities might be different, a question which is accessible in numerical investigations.

Section II B proposes that the $t^\beta$ increase of the relaxational response, at times which are shorter than the terminal time $\tau_c$, is due to Eshelby transformations of regions with increasing size. The Kohlrausch $\beta$ is then given by the ratio of the structural entropy increase per particle to the barrier increase per particle. This allows to calculate the number of atoms or molecules of the Eshelby region with the terminal relaxation, at the crossover from reversible to irreversible Eshelby transformations.

Section III B calculates values for twelve substances, among them two metallic glasses. As Section III C shows, the proposal finds experimental support from shear relaxation data in a metallic glass at room temperature \cite{atzmon} and in silica over the whole temperature range from low temperatures to $T_g$ \cite{marx}, from which one gets Eshelby region sizes close to the prediction.

In addition, Section III C corroborates a finding of Paper II in silica \cite{asyth1}, namely a Kohlrausch barrier density extending with a constant Kohlrausch $\beta$ down to less than half the glass transition barrier $V_c$. Section III C shows the same extension of the Kohlrausch barrier density down to low barriers for two more examples, namely the mentioned metallic glass and selenium. This disproves a pet idea of the author, namely that the Kohlrausch $\beta$ is due to the interaction of the Eshelby states, an idea on which he wrote several papers, which now appear irrelevant. The idea must be wrong, if the Kohlrausch behavior continues in an unchanged way down into a range where one can reckon with only a few isolated double well potentials.

But the new Adam-Gibbs explanation of the Kohlrausch law is not without question marks. One of them is the remarkable constancy of $\beta$ as a function of temperature up to rather high temperatures \cite{met}, which is also seen in Table I of Paper II \cite{asyth1} and Fig. 6 of reference \cite{albena}. If $\beta=S_1T/V_1$, as postulated in eq. (\ref{beta}), $\beta$ should increase with increasing temperature, because $S_1$ increases even more than $T$ itself and $V_1$ should rather decrease. Also, the undeniable tendency \cite{albena} to the value 1/2 remains unexplained.

One gets a hint for the reason behind these inconsistencies by looking at the values of the average squared jump width $\overline{e^2}$ of about forty to fifty in Table II for the metallic glasses. In metallic glasses, the average product $GV_a/k_BT_g$ has the value 17.6 from many measurements at the glass transition \cite{john}. In Section II C, two mechanisms for the highly viscous flow of simple liquids have been discussed. The first was the motion of three neighbors suggested by the string motion \cite{olig,glotzer} for $N_1=3$ atoms and an $\epsilon_1=0.292$, yielding a value $\overline{e^2}=1.1$ from eq. (\ref{eq}), obviously much too small to explain the experimental finding in Table II. The other is the gliding triangle motion \cite{buscho} with $N_1=6$ and $\epsilon_1=1/\sqrt{2}$, which provides $\overline{e^2}=13.2$. This second value is better, but still a factor of three too small.

Similar results are obtained for the molecular glass formers. Taking the example of m-toluidine, a rigid molecule, one calculates $GV_a/k_BT_g=75.9$ from the fit of $G(\omega)$ in Table I of Paper II \cite{asyth1}. With $N_1=2$ molecules, one requires an $\epsilon_1^2=2.3$ to get the observed value for $\overline{e^2}$, a bit too large to be believed, because it corresponds to a shear angle of 90 degrees. A local shear angle of 60 degrees for the two molecules seems much more probable, the same discrepancy as in the metallic glasses.

The most probable explanation for these discrepancies lies in the assumption of completely uncorrelated local uniform strain distortions in the structural arrangement of Section II C. The value of $\overline{e^2}$ would be very susceptible even to a small amount of correlation. If the correlation increased with decreasing temperature, it could compensate the decrease of the entropy and lead to a temperature-independent Kohlrausch $\beta$. But at our present stage of understanding, this is just a speculation.    

To conclude, the description of the highly viscous flow in terms of irreversible Eshelby transitions has been further detailed, predicting different viscosities in stationary and alternating flow and proposing an Adam-Gibbs explanation for the Kohlrausch $\beta$. The results compare favorably with numerous experimental data from the literature.


\begin{thebibliography}{99}
\bibitem{bnap} R. B\"ohmer, K. L. Ngai, C. A. Angell, and D. J. Plazek, J. Chem. Phys. {\bf 99}, 4201 (1993)
\bibitem{greg} G. B. McKenna, J. Non-Cryst. Solids {\bf 172-174}, 756 (1994)
\bibitem{nemilov} S. V. Nemilov,{\it Thermodynamic and Kinetic Aspects of the Vitreous State} (CRC, Boca
Raton, FL 1995)
\bibitem{angell} C. A. Angell, J. Res. Natl. Inst. Stand. Technol. {\bf 102}, 171 (1997) 
\bibitem{dyre} J. C. Dyre, Rev. Mod. Phys. {\bf 78}, 953 (2006)
\bibitem{cavagna} A. Cavagna, Phys. Rep. {\bf 476}, 51 (2009)
\bibitem{tinaa} T. Hecksher, N. B. Olsen, K. Niss, and J. C. Dyre, J. Chem. Phys. {\bf 133}, 174514 (2010)
\bibitem{bb} L. Berthier and G. Biroli, Rev. Mod. Phys. {\bf 83}, 587 (2011)
\bibitem{ngai} K. L. Ngai, {\it Relaxation and Diffusion in Complex Systems}, (Springer, New York 2011)
\bibitem{shov2015} T. Hecksher and J. C. Dyre, J. Non-Cryst. Solids {\bf 407}, 14 (2015)
\bibitem{royall} C. P. Royall and S. R. Williams, Phys. Rep. {\bf 560}, 1 (2015)
\bibitem{lerner} E. Lerner and E. Bouchbinder, J. Chem. Phys. {\bf 148}, 214502 (2018)
\bibitem{albena} A. I. Nielsen, T. Christensen, B. Jakobsen, K. Niss, N. B. Olsen, R. Richert, and J. C. Dyre, J. Chem. Phys. {\bf 130}, 154508 (2009)
\bibitem{asyth} U. Buchenau, J. Chem. Phys. {\bf 148}, 064502 (2018)
\bibitem{asyth1} U. Buchenau, J. Chem. Phys. {\bf 149}, 044508 (2018)
\bibitem{eshelby} J. D. Eshelby, Proc. Roy. Soc. {\bf A241}, 376 (1957)
\bibitem{adam} G. Adam and J. H. Gibbs, J. Chem. Phys. {\bf 43}, 139 (1958)
\bibitem{olig} H. R. Schober, C. Oligschleger, and B. B. Laird, J. Non-Cryst. Solids {\bf 156}, 965 (1993); C. Oligschleger and H. R. Schober, Phys. Rev. B {\bf 59}, 811 (1999)
\bibitem{glotzer} C. Donati, J. F. Douglas, W. Kob, S. J. Plimpton, P. H. Poole, and S. C. Glotzer, Phys. Rev. Lett. {\bf 80}, 2338 (1998)
\bibitem{buscho} U. Buchenau and H. R. Schober, Phil. Mag. B {\bf 88}, 3385 (2008)
\bibitem{helfand} E. Helfand, J. Polym. Sci., Polym. Symp. {\bf 73}, 39 (1985)
\bibitem{tnb} D. J. Plazek and J. H. Magill, J. Chem. Phys. {\bf 45}, 3038 (1966)
\bibitem{bero} D. J. Plazek, C. A. Bero and I.-C. Chay, J. Non-Cryst. Solids {\bf 172-174}, 181 (1994)
\bibitem{tnb2} D. J. Plazek, J. H. Magill, I. Echeverria, and I.-C. Chay, J. Chem. Phys. {\bf 110}, 10445 (1999)
\bibitem{se} W. A. Phillips, U. Buchenau, N. N\"ucker, A. J. Dianoux, and W. Petry, Phys. Rev. Lett. {\bf 63}, 2381 (1989)
\bibitem{tina} T. Hecksher, N. B. Olsen, K. A. Nelson, J. C. Dyre and T. Christensen, J. Chem. Phys. {\bf 138}, 12A543 (2013)
\bibitem{niss} B. Jakobsen, K. Niss, and N. B. Olsen, J. Chem. Phys. {\bf 123}, 234510 (2005)
\bibitem{ra}  R. Richert and C. A. Angell, J. Chem. Phys. {\bf 108}, 9016 (1998)
\bibitem{chang} S. S. Chang and A. B. Bestul, J. Chem. Thermodynamics {\bf 6}, 325 (1974)
\bibitem{wildek} G. Wilde, G. P. G\"orler, R. Willnecker, and G. Dietz, Appl. Phys. Lett. {\bf 65}, 397 (1994)
\bibitem{busch} R. Busch, A. Masuhr, and W.L. Johnson, Mat. Sci. Eng. A {\bf 304–306}, 97 (2001)
\bibitem{nem} S. V. Nemilov, V. N. Bogdanov, A. M. Nikonov, S. N. Smerdin, A. I. Nedbai and B. F. Borisov, Fiz. i Khim. Stekla {\bf 13}, 801 (1987) (Sov. J. Glass Phys. Chem. {\bf 13}, 413 (1987))
\bibitem{gundermann} D. Gundermann, U. R. Pedersen, T. Hecksher, N. P. Bailey, B. Jakobsen, T. Christensen, N. B. Olsen, T. B. Schroeder, D. Fragiadakis, R. Casalini, C. M. Roland, J. C. Dyre and K. Niss, Nature Physics {\bf 7}, 816 (2011)
\bibitem{cpppe} B. Jakobsen, N. B. Olsen, and T. Christensen, Phys. Rev. E {\bf 81}, 065505 (2010)
\bibitem{cptol} G. Pratesi, P. Bartolini, D. Sinatra, M. Ricci, R. Righini, F. Barocchi, and R. Torre, Phys. Rev. E {\bf 67}, 021505 (2003)
\bibitem{cppc} H. Fujimori and M. Oguni, J. Chem. Thermodynamics {\bf 26}, 367 (1994)
\bibitem{wunderlich} B. Wunderlich, J. Chem. Phys. {\bf 64}, 1052 (1960)
\bibitem{atzmon} J. D. Ju, D. Jang, A. Nwankpa, and M. Atzmon, J. Appl. Phys. {\bf 109}, 053522 (2011)
\bibitem{alni} P. Si, X. Bian, J. Zhang, H. Li, M. Sun, and Y. Zhao, J. Phys.: Condens. Matter {\bf 15}, 5409 (2003)
\bibitem{burel} U. Buchenau, Phys. Rev. B {\bf 63}, 104203 (2001)
\bibitem{gilroy} K. S. Gilroy and W. A. Phillips, Phil. Mag. B {\bf 43}, 735 (1981)
\bibitem{philmag2002} U. Buchenau, A. Wischnewski, R. Zorn, and N. Hadjichristides, Phil. Mag. B {\bf 82}, 209 (2002)
\bibitem{marx} J. W. Marx and J. M. Sivertsen, J. Appl. Phys. {\bf 24}, 81 (1953)
\bibitem{kirby} P. L. Kirby, J. Soc. Glass Technol. {\bf 34}, 388 (1954)
\bibitem{pino} G. Carini Jr., M. Cutroni, G. Galli, and F. Wanderlingh, J. Non-Cryst. Solids {\bf 30}, 61 (1978)
\bibitem{liu} X. Liu, T. H. Metcalf, M. R. Abernathy, and R. B. Stephen, Mat. Res. {\bf 21} (suppl. 2): e20170881 (2018)
\bibitem{roland} C. M. Roland, P. G. Santangelo, D. J. Plazek, and K. M. Bernatz, J. Chem. Phys. {\bf 111}, 9337 (1999)
\bibitem{galli} G. Galli, P. Migliardo, R. Bellisent, and W. Reichardt, Solid State Commun. {\bf 57}, 195 (1986)
\bibitem{kozh} V.F. Kozhevnikov, W.B. Payne, J.K. Olson, A. Allen, and P.C. Taylor, J. Non-Cryst. Solids {\bf 353}, 3254 (2007) 
\bibitem{met} L.-M. Wang, R. Liu, and W. H. Wang, J. Chem. Phys. {\bf 128}, 164503 (2008)
\bibitem{john} W.L. Johnson and K. Samwer, Phys. Rev. Lett. {\bf 95}, 195501 (2005)
\end{thebibliography}
\end{document}